\documentclass[preprint,authoryear,12pt]{elsarticle}

\usepackage{epsfig}
\usepackage{verbatim}
\usepackage{amssymb}
\usepackage{ulem}

\usepackage[ps2pdf,%
a4paper=true,%
breaklinks=true,%
colorlinks=true,%
pdfauthor={Kravchenko et al.},%
pdftitle={The jet of S5~0716$+$71 at $\mu$as scales with \textit{RadioAstron}}%
]{hyperref}

\journal{Advances in Space Research}

\begin{document}

\begin{frontmatter}



\title{The jet of S5~0716$+$71 at $\mu$as scales with \textit{RadioAstron}} 


\author{Evgeniya V. Kravchenko\corref{cor}}
\address{INAF Istituto di Radioastronomia, Via P. Gobetti, 101, Bologna, 40129, Italy;\\Astro Space Center, Lebedev Physical Institute, Russian Academy of Sciences, Profsouznaya st., 84/32, Moscow, 117997, Russia}
\cortext[cor]{Corresponding author}
\ead{e.kravchenko@ira.inaf.it}


\author{Jos\'e L. G\'omez\corref{}}
\address{Instituto de Astrof\'isica de Andaluc\'ia, CSIC, Glorieta de la Astronom\'ia s/n, Granada, 18008, Spain}
\ead{jlgomez@iaa.es}

\author{Yuri Y. Kovalev\corref{}}
\address{Astro Space Center, Lebedev Physical Institute, Russian Academy of Sciences, 
Profsouznaya st., 84/32, Moscow, 117997, Russia\\Moscow Institute of Physics and Technology, Dolgoprudny, Institutsky per., 9, Moscow region, 141700, Russia\\Max-Planck-Institut f\"{u}r Radioastronomie, Auf dem H\"{u}gel 69, Bonn, 53121, Germany}
\ead{yyk@asc.rssi.ru}

\author{Petr A. Voytsik\corref{}}
\address{Astro Space Center, Lebedev Physical Institute, Russian Academy of Sciences, 
Profsouznaya st., 84/32, Moscow, 117997, Russia}

\author{on behalf of the \textit{RadioAstron} Polarization KSP and AGN Survey KSP\corref{}}

\begin{abstract}
Ground-space interferometer \textit{RadioAstron} provides unique opportunity to probe detail structure of the distant active galactic nuclei at $\mu$as scales.
Here we report on \textit{RadioAstron} observations of the BL~Lac object S5~0716$+$71, performed in a framework of the AGN Polarization and Survey Key Science Programs at 22~GHz during 2012-2018.
We obtained the highest angular resolution image of the source to date, at $57\times24~\mu$as. 
It reveals complex structure of the blazar jet in the inner 100~$\mu$as, with emission regions that can be responsible for the blazar variability at timescales of a few days to week. Linear polarization is detected in the core and jet areas at the projected baselines up to about $5.6$~Earth diameters. The observed core brightness temperature in the source frame of $\geq2.2\times10^{13}$~K is in excess of theoretical limits, suggesting the physical conditions are far from the equipartition between relativistic particles and magnetic field.
\end{abstract}

\end{frontmatter}

\parindent=0.5 cm

\section{Introduction}

The international space VLBI mission \textit{RadioAstron} \citep{2017SoSyR..51..535K} features a 10-m radio telescope (SRT) onboard of SPEKTR-R spacecraft, delivering the highest angular resolution to date ($\thicksim$7$\mu$as at 1.3~cm, see Kovalev et al. these proceedings).
One of the \textit{RadioAstron} Key Science Programs (KSP) focuses on polarimetric studies of the most active and highly polarized active galactic nuclei (AGN) in the sky.
The instrumental polarization of the SRT is less then 9\% \citep{2015A&A...583A.100L,2016ApJ...817...96G}, confirming the \textit{RadioAstron} capabilities for polarization imaging.
During more than five years, the polarization KSP has already delivered the most detailed study of a number of AGN, e.g. TXS 0642$+$449 \citep{2015A&A...583A.100L}, BL~Lacertae \citep{2016ApJ...817...96G} and 3C~273 \citep{2017A&A...604A.111B}.
Another one of the most relevant \textit{RadioAstron} Key Science Program is the Survey and Monitoring of AGNs, which aims at investigating the extremely high brightness temperature of AGNs with baselines reaching $\thicksim28$ Earth diameters (ED) (see Kovalev et al. these proceedings for further details on the program).

BL~Lacertae object S5~0716$+$71 (hereafter blazar 0716$+$714), is a key target for both of these KSPs.
0716$+$714 is famous for its extreme variability across the electromagnetic spectrum, including intraday variability (IDV).
The IDV phenomenon, after its discovery in the middle of 80s \citep{1986MitAG..65..239W,1987AJ.....94.1493H}, still remains a matter of debate.
IDV has been commonly observed in compact flat-spectrum radio sources \citep[e.g.][]{2003AJ....126.1699L} and appears to be correlated  with the compactness of the VLBI core \cite[e.g.][]{1997ApJ...490L...9K}.
Considerable evidence has accumulated that interstellar scintillation is the principal mechanism responsible for IDV at cm and shorter wavelengths \citep[ISS; e.g.][and references therein]{2018MNRAS.474.4396K}.

On the other hand, there are evidence that IDV is being produced by processes intrinsic to the relativistic jet, like frequency dependence of the IDV amplitude, highly polarized micro-flares, multi-frequency correlation and others.

Blazar 0716$+$714 is considered as one of the best candidates for having an intrinsic origin of the observed IDV, since there are some properties that cannot be explained by ISS: the correlation between optical brightness and radio spectral index, as well as simultaneous change in variability time-scale during observations in 1990 \citep{1990A&A...235L...1W,1991ApJ...372L..71Q, 1996AJ....111.2187W}, observed increase in variability amplitude with frequency \citep{2008A&A...490.1019F}, rapid variability at millimeter wavelengths \citep{2006A&A...451..797O}, quasi-periodic intrahour oscillations at optical band \citep{2010ApJ...719L.153R}, and highly polarized optical micro flares \citep{2015ApJ...809L..27B}.
However, the observed annual modulation \citep{2012A&A...543A..78L} implies that some of the  0716$+$714 variability is indeed produced by interstellar scintillation, at least at 6~cm and 11~cm \citep{2012MNRAS.425.1357G}.
The motivation of this study was to probe the jet structure at the finest angular resolution and to investigate the origin of its IDV.
Further details and other results of this work will be presented in Kravchenko et al. (ApJ in prep.) and Kravchenko et al. (A\&A in prep.).

\section{Obsrevations}

Observations of the blazar as part of the Polarization KSP has been performed on 2015 January 3-4 during 12-hour global space VLBI session at 22~GHz.
The signal has been recorded in two circular polarizations (right- and left-hand) in a bandwidth of 32~MHz for the space and 64~MHz for the ground antennas.
An array of 11 ground antennas and SRT tracked the source: Brewster, Effelsberg, Ford Davis, Green Bank, Hancock, Los Alamos, Noto, Owens Valley, Pie Town, Shangai and Torun.
Fringes between ground antennas and space telescope has been found at the maximum projected baseline of 5.25~G$\lambda$ (5.56 ED or 70,833~km).

The AGN Survey KSP consists on snapshot-mode ($\leq1$hour long) observations of the sources with a number of antennas, depending on the availability at the time.
0716$+$714 has been observed 68 times within this program between 2012 and 2018 at 1.7, 4.8 and 22~GHz. 
The maximum projected baseline in these observations reached about 25.5 ED (324,900~km), during the segment taken on 2013 November 7 at 1.7~GHz, when blazar was successfully detected.

\section{Results and discussion}

\begin{figure}
\begin{center}
\includegraphics[width=6.6cm,angle=-90]{pgplot.ps}\quad
\includegraphics[width=7.0cm,angle=-90]{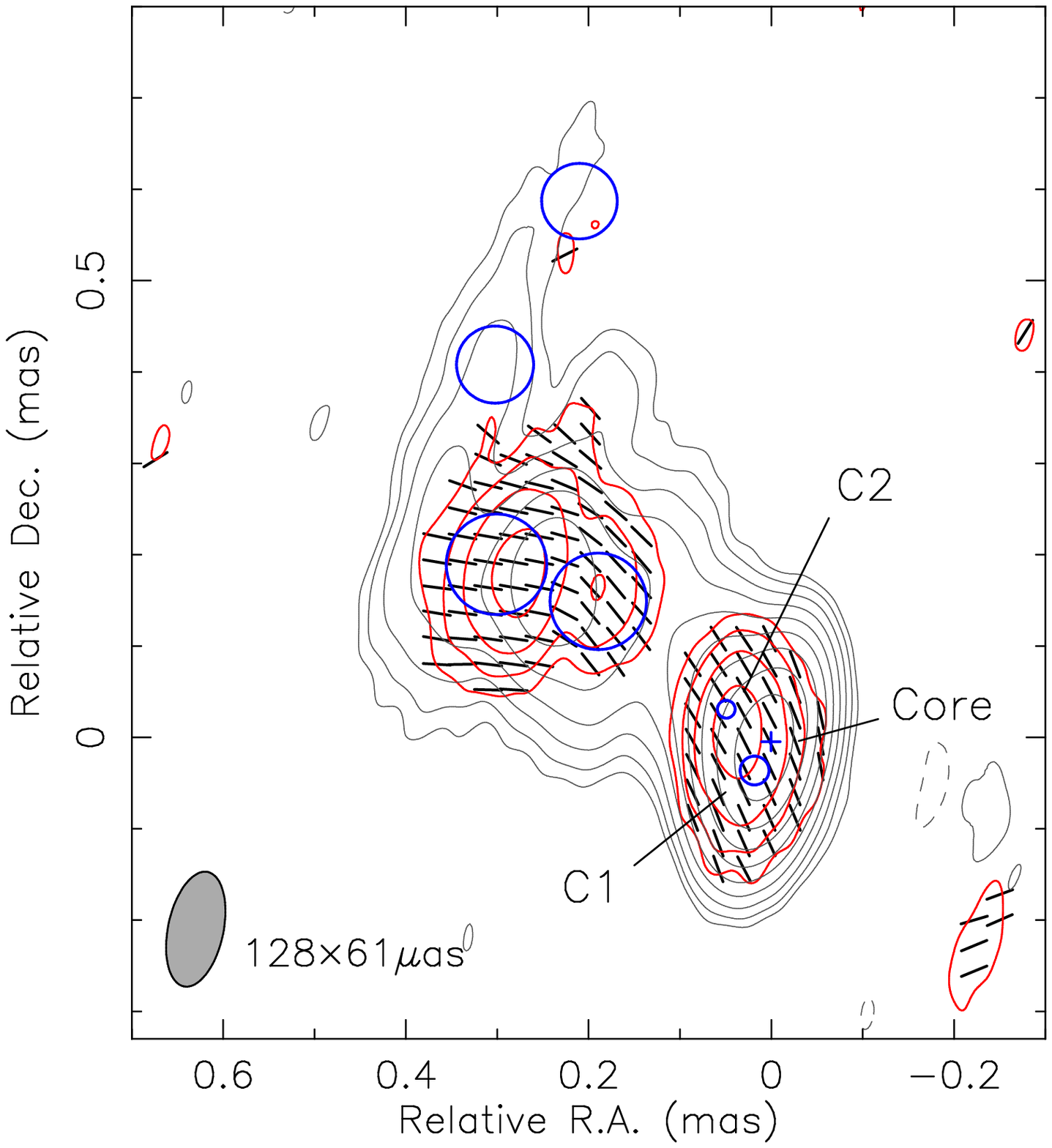}
\end{center}
\caption{(\textit{uv})-coverage (left) and polarimetric \textit{RadioAstron} image of 0716$+$714 (right) made on 2015 January 3-4 at 22~GHz, using uniform weighting. The corresponding synthesized beam is displayed by shaded ellipse and is given at $-11^{\circ}$. Red solid lines mark linearly polarized contours, black solid lines indicate orientation of the electric vector position angle, uncorrected for the rotation measure. Model fitted components for the jet region are given by blue circles and core component is given by cross. The lower contour for the total intensity is shown at 1.4~mJy/beam, for the polarized intensity at 1.5~mJy/beam.\label{figure1}}
\end{figure}

The polarization space VLBI image of the blazar is shown in Fig.~\ref{figure1}, using uniform weighting at an angular resolution of $\thicksim61\mu$as.
Giving more weights to the \textit{RadioAstron} data, thus assuming super uniform weighting scheme with a binning size of 2 (\textit{uv})-grid pixels and no weighting of the visibilities by the errors, the maximum angular resolution of the imaging experiment then reaches {$\thicksim24~\mu$as}.
Zooming in at this highest angular resolution, 0716$+$714 image shows a complex bent structure in the central $\sim100~\mu$as core region, consisting on an unresolved core and nearby components C1 and C2, located at $\thicksim41~\mu$as and $\thicksim58~\mu$as from the core, respectively. 
The jet initially extends towards the south-east (C1), at a position angle of $153^{\circ}$, followed by a sharp bending of about $95^{\circ}$ towards the north-east (C2), maintaining that direction for about 1~mas until another sharp bend towards the north-west is observed.
Such extreme orientation of the central $100-200~\mu$as region relative to the position angle of the jet at larger scales in the blazar is not surprising: \citet{2015A&A...578A.123R, 2014A&A...571L...2R} and \citet{2017ApJ...846...98J} have already observed nearly $90^{\circ}$ offset in the orientation between these two jet regions .

We estimate the size of the core to be $\leq12\times5~\mu$as, while the sizes of components C1 and C2 are $\thicksim32~\mu$as and $\thicksim19~\mu$as, respectively.
Using model fitted parameters and following \citet{2005AJ....130.2473K}, we obtain the observed brightness temperatures in the rest frame of the source of $(6.99\pm0.18)\times10^{11}$~K for C1, $(1.20\pm0.06)\times10^{12}$~K for C2, and $T_\mathrm{b}\geq2.2\times10^{13}$~K for the core. 

\begin{figure}
\begin{center}
\includegraphics[width=5.85cm]{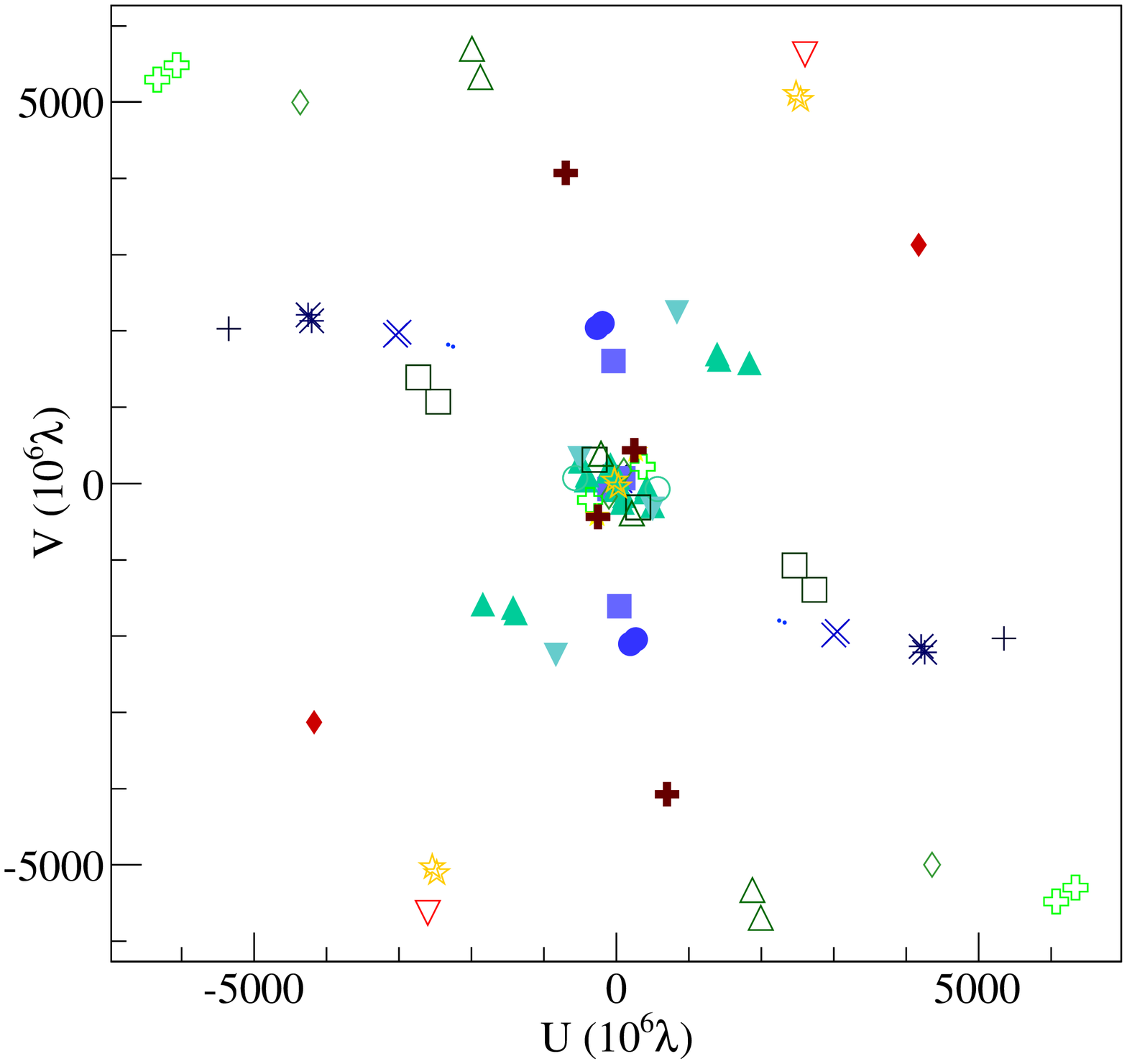}\quad
\includegraphics[width=7.35cm]{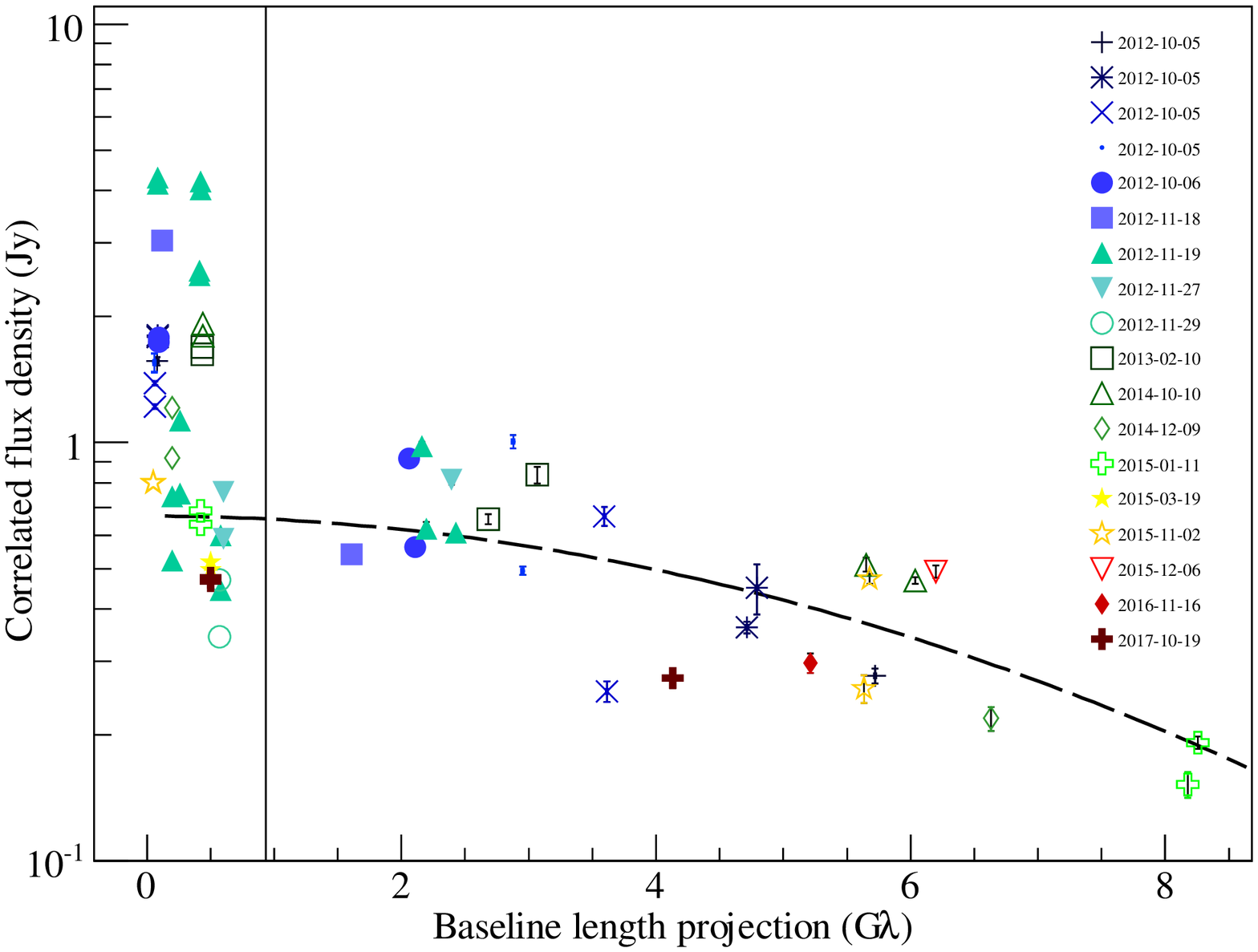}
\end{center}
\caption{\textit(uv)-coverage (left) and visibility amplitudes versus \textit{uv}-distance (right) of the \textit{RadioAstron} observations of 0716$+$714 at 22~GHz in 2012--2017. Solid vertical line marks the equivalent Earth diameter. Dashed line represents the model fit results by Gaussian distribution centered at zero baseline.\label{figure2}}
\end{figure}

The limited $(u,v)$-coverage of an individual \textit{RadioAstron} snapshot-mode observations taken within AGN Survey KSP  does not allow for imaging and model fitting the brightness distribution. Whereas blazar has been targeting 18 times at 22~GHz during 2012--2017 within the program, we combine these epochs and consider them together. (\textit{uv})-coverage for these observations is given in Fig.~\ref{figure2}, along with visibility distribution versus (\textit{uv})-distance.

Assuming that interferometric visibility at space-ground baselines is determined by the most compact component (see \cite{2015A&A...574A..84L} for discussion), its size can be estimated. We represent its contribution $V$ by a circular Gaussian function considering only space-ground baselines $l$ in a form $V = A\mathrm{exp}(-l^2/(2C^2))$, where $A$ and $C$ are the fitted parameters.
The size $\theta$ of the corresponding emitting region is then derived as $\theta=\sqrt{2\mathrm{ln}2}\times206264.806 / (\pi C)$ arcseconds
which results in $\thicksim15~\mu$as, and the brightness temperature of $\thicksim10^{13}$~K.
One need to assume, that these values are a subject of extreme blazar variability and a more complex structure of the jet in the central region, rather than a single compact component. How large is the contribution of these factors is the matter of future studies, meanwhile the values of $\theta$ and $T_\mathrm{b}$, obtained from the AGN Survey KSP, are close to the corresponding values, coming from the \textit{RadioAstron} imaging experiment.

In the case of inverse Compton losses for incoherent synchrotron sources, it is expected that the intrinsic brightness temperature does not exceed $\sim10^{11.5}$~K \citep{1969ApJ...155L..71K}.
In case of equipartition between the energies of the magnetic field and radiating particles, an upper limit of $T_\mathrm{b,int}\leq5\times10^{10}$~K \citep{1994ApJ...426...51R} is expected.
The intrinsic and observed brightness temperatures are connected through the Doppler boosting factor as $T_\mathrm{b,int}=T_\mathrm{b,obs}/\delta$.
Considering a Doppler factor of $\delta\thicksim24$, the maximum observed in the blazar \citep[e.g.][]{2015A&A...578A.123R, 2017ApJ...846...98J, 2018ApJ...866..137L}, the intrinsic brightness temperature of the core results in $\gtrsim9\times10^{11}$~K and is in excess of Compton limit and equipartition brightness temperature value. 
This result supplements recent \textit{RadioAstron} studies of a quasar 3C~273 \citep{2016ApJ...820L...9K} and a blazar BL~Lac \citep{2016ApJ...817...96G}, where authors also found that $T_\mathrm{b,int}$ breaks the abovementioned limits.
Excess of $T_\mathrm{b,int}$ over the synchrotron limit could indicate that some other mechanisms are at work, such as coherent radiation mechanism \citep{1992ApJ...391L..59B}, quasi-monoenergetic electron population \citep{2007A&A...463..145T} or proton synchrotron radiation \citep{2000ARep...44..719K}.


The radio polarimetric structure of the blazar at the highest resolution is characterized by 15\%-linearly polarized compact component, located $\sim58~\mu$as down stream the core (see Fig.~\ref{figure1}).
The position of this polarized feature coincides with the location of model fitted component at $58~\mu$as.
Considering that Doppler-adjusted light-crossing time across jet component governs the observed timescale, the variability timescale can be estimated as $\tau (\mathrm{yr}) = \big[25.3(R/\mathrm{mas})(D_\mathrm{L}/\mathrm{Gpc})\big]/\big[\delta(1+z)\big]$ \citep{2017ApJ...846...98J}, where $D_\mathrm{L}$ is the luminosity distance, which equals to 1612.6~Mpc (assuming flat $\Lambda$CDM cosmology with $\Omega_\mathrm{m}=0.3$, $\Omega_{\Lambda}=0.3$ and $H_0=70$~km~s$^{-1}$Mpc$^{-1}$) at the redshift of 0.31 \citep{2013ApJ...764...57D, 2008A&A...487L..29N}, and $R$ is the angular size of the model fitted component.
For the polarized component C2, with an angular size of $19~\mu$as, $\tau$ is about 9 days, meanwhile for the unresolved core it is of $\thicksim2.5$ days.
Thus, intrinsic processes in the blazar jet can explain its variability on the time scales of a days to week.
To explain five-hour-long, $\leq60$\% optical polarized micro flare observed in the blazar \citep{2015ApJ...809L..27B}, even smaller regions with a highly ordered magnetic field are required.

Linear polarization structure of 0716$+$714 jet, given in Fig.~\ref{figure1}, is dominated by two features, associated with the core region and with the jet emission. Uncorrected for the Faraday rotation, electric vector position angle (EVPA) follows the orientation of the jet for both these polarized regions.
\citet{2016A&A...592L..10L} in their study measured RM between 22~GHz and 43~GHz ranging between $-9200$ to 6300 rad/m$^2$, and $-71000$ to $-7300$ rad/m$^2$ between 43~GHz and 86~GHz. Assuming that the rotation measure value in the central core region is negative and of a few thousands, then the orientation of the EVPA will be better aligned with the core-C1 component, indicating dominance of the toroidal component of the ordered magnetic field in the blazar jet.

\vspace{1cm}

Our results confirm the powerful capabilities of the \textit{RadioAstron} mission to probe relativistic jets in vicinity of the central black hole, including high angular resolution polarization imaging.
Our imaging results from 2015 January 3-4 gave an angular resolution of $24\mu$as, a factor of two higher than was achieved with ground VLBI arrays to date \citep[e.g.][]{2016JPhCS.718e2032R}.


\section{Acknowledgments}
EVK acknowledges support from the Italian Space Agency under contract ASI-INAF 2015-023-R.O.
JLG was supported by the Spanish Ministry of Economy and Competitiveness grants AYA2013-40825-P and AYA2016-80889-P.
YYK was partly supported by the Russian Foundation for Basic Research (project 17-02-00197). The RadioAstron project is led by the Astro Space Center of the Lebedev Physical Institute of the Russian Academy of Sciences and the Lavochkin Scientific and Production Association under a contract with the State Space Corporation ROSCOSMOS, in collaboration with partner organizations in Russia and other countries. The European VLBI Network is a joint facility of independent European, African, Asian, and North American radio astronomy institutes. Scientific results from data presented in this publication are derived from the following EVN project code: GL041. The National Radio Astronomy Observatory is a facility of the National Science Foundation operated under cooperative agreement by Associated Universities, Inc.



\end{document}